\documentclass[journal]{IEEEtran}
\usepackage{amsmath,amsfonts}
\usepackage{algorithmic}
\usepackage{array}
\usepackage[caption=false,font=normalsize,labelfont=sf,textfont=sf]{subfig}
\usepackage{textcomp}
\usepackage{stfloats}
\usepackage{url}
\usepackage{verbatim}
\usepackage{graphicx}
\def\BibTeX{{\rm B\kern-.05em{\sc i\kern-.025em b}\kern-.08em
    T\kern-.1667em\lower.7ex\hbox{E}\kern-.125emX}}
\usepackage{balance}
\usepackage{threeparttable, booktabs, ragged2e, dcolumn}

\begin{document}
\title{In-the-wild vibrotactile sensation: Perceptual transformation of vibrations from smartphones}
\author{
    \IEEEauthorblockN{Keiko Yamaguchi\IEEEauthorrefmark{1}, Satoshi Takahashi\IEEEauthorrefmark{2}, \textit{Member, IEEE}}
    \\\IEEEauthorblockA{\IEEEauthorrefmark{1}Department of Economics, Nagoya University
    \\keiko.yamaguchi@soec.nagoya-u.ac.jp}
    \\\IEEEauthorblockA{\IEEEauthorrefmark{2}College of Science and Engineering, Kanto Gakuin University
    \\satotaka@kanto-gakuin.ac.jp}
    \thanks{This study was supported by JSPS Grant-in-Aid for Early-Career Scientists (KAKENHI) Grant Number 20K13615.}
}


\maketitle

\begin{abstract}

Vibrations emitted by smartphones have become a part of our daily lives.
The vibrations can add various meanings to the information people obtain from the screen.
Hence, it is worth understanding the perceptual transformation of vibration with ordinary devices to evaluate the possibility of enriched vibrotactile communication via smartphones.
This study assessed the reproducibility of vibrotactile sensations via smartphone in the in-the-wild environment.
To realize improved haptic design to communicate with smartphone users smoothly, we also focused on the moderation effects of the in-the-wild environments on the vibrotactile sensations: the physical specifications of mobile devices, the manner of device operation by users, and the personal traits of the users about the desire for touch.
We conducted a Web-based in-the-wild experiment instead of a laboratory experiment to reproduce an environment as close to the daily lives of users as possible.
Through a series of analyses, we revealed that users perceive the weight of vibration stimuli to be higher in sensation magnitude than intensity under identical conditions of vibration stimuli.
We also showed that it is desirable to consider the moderation effects of the in-the-wild environments for realizing better tactile system design to maximize the impact of vibrotactile stimuli.

\end{abstract}


\section{Introduction}
\noindent
The sensation of vibrotactile stimuli has been studied for a long time \cite{Verrillo1969-kl}, and extensive research has clarified the type of sensations that can be designed using different vibration actuators \cite{Chang2002-kh},\cite{Choi2013-zs}.
Recently vibrotactile perceptions under multimodality have garnered research interest \cite{hoggan2008crossmodal},\cite{Rantala2020-ns}.

Smartphones have now become a necessity in our lives and are one of the most familiar vibration actuators.
Vibrations emitted by smartphones can impart various meanings to the information that users obtain from the screen \cite{Rantala2020-ns}.
With the widespread use of smartphones, vibratory stimuli have become a part of our daily lives \cite{Blum2019-ud}; the usage environment of smartphones is entirely different from the ones assumed in previous studies.
Hence, it is worth understanding how people perceive the vibrotactile stimuli from smartphones for better information transmission in everyday use.

This study focuses on vibrotactile sensations generated from smartphones under real-life constraints.
We assessed the reproducibility of vibrotactile sensations via in-the-wild experiment.
We investigated the perceptual transformation of vibrations using ordinary devices to evaluate the possibility of enriched vibrotactile communication on smartphones.
We also examined the moderation effects of the in-the-wild environments on the vibrotactile sensations for realizing better haptic designs for smooth communicate for smartphone users.

\section{Related works and motivation}
\noindent
Vibration is a stimulus that can elicit multiple tactile perceptions.
Previous studies elucidated vibrotactile percepts such as roughness and smoothness that can be differentiated by the vibration frequency \cite{tan1999information},\cite{park2011perceptual}, a virtual texture of materials reproduced on a touchscreen \cite{romano2011creating}, and softness that requires consideration of the material of the vibration actuator \cite{Visell2014}.
Most of these studies focused on designing apparatuses and vibration patterns to elicit certain perceptions in experimental settings \cite{Blum2019-ud}.
Few studies have clarified how humans perceive vibration stimuli from actuators with specific physical properties under in-the-wild conditions.

Vibration stimuli from smartphones are mainly used to convey immaterial information, such as incoming call notifications and task completion.
For this purpose, the intensity of stimuli plays an essential role in communicating critical and emergent information to reduce the subjective workload of users \cite{hoggan2008investigating},\cite{park2011tactile},\cite{Palomares2020-mw}.
Currently, people possess digital assets such as digital wallets on their smartphones. 
Since digital assets have no substance, it is difficult to perceive their existence.
Weight is an essential factor in the perception of an entity.
Researchers utilize vibrotactile feedback for object recognition on touch screens and in virtual spaces \cite{Park2020-db},\cite{Khosravi2022-ze}.
It is reasonable to assume that the perceived weight of digital assets can enhance the perception of their substance; this is an attractive hypothesis owing to its commercial impact \cite{Choi2013-zs}.
Some studies mapped smartphone vibration perception to the psychological state of users \cite{Manshad2019-gv},\cite{Palomares2020-mw}, but few directly mapped it to the tactile perception.
Hence, in this study, we focus on the perceived intensity and weight of vibrotactile stimuli from smartphones.

We selected an iPhone as an apparatus in this study because it is one of the most popular vibration actuators today and the same haptic engine, i.e., Core Haptics \cite{core-haptics}, is employed in iPhone 8 and later models.
We conducted a Web-based in-the-wild experiment \cite{Abou_Chahine2022-vj} instead of a laboratory experiment.
In our experiment, participants worked on assignments in an environment that was as close as possible to their daily lives.

\section{Research questions}
\noindent
The primary research question of this study was how people perceived and rendered the vibrations of their smartphones as vibrotactile stimuli and evaluated the magnitude of vibrotactile sensations: intensity and weight.
The intensity of vibrotactile stimuli is a straightforward sensation that can be evaluated by ordinary smartphone users.
Meanwhile, the weight of vibration is a more complicated perception than the intensity: it is a perceptual transformation that may occur in the mind of the user.
Hence, the following research question was derived:

\medskip
\noindent
\textbf{RQ1}: Do people evaluate different qualitative sensations of the same vibrotactile stimuli on the smartphone similarly or differently?

\medskip
In our daily environment, external and internal factors affect how people perceive vibrotactile stimuli: the physical specifications of devices, the manner of their use, and the personal traits of users.
Previous studies showed that the physical specifications of devices influence the qualities of the vibrotactile stimuli \cite{johansson2009coding}.
The iPhone has a simple rectangular, common form unlike the original devices developed in the previous studies; however, the sizes and weights of the models vary.

In addition, the manner of operation of the devices affects vibrotactile perception \cite{johansson2009coding}: in particular, the manner of holding the iPhone.
We focused on which hand people usually use to hold the iPhone whether they use the dominant hand or not. How the brains react to vibrotactile stimuli regarding hand dominance has been studied in the field of neuroscience \cite{Yang2018-gy}, \cite{Kim2020-do}; however, we focused on how people perceived the stimuli and evaluated them as interpretable sensations for the application of implications to design vibrotactile stimuli.

Finally, it was necessary to consider the personal traits of smartphone users, such as the desire for touch when extracting information, when examining how they perceive smartphone vibrations.
The ``Need for Touch'' (NFT) scale \cite{Peck2003-bi} indicates individual differences in preference for haptic (touch) information.
People with higher NFT scores are likely to wish to touch objects to extract precise information about them or to enjoy haptic perceptions from them.
We hypothesized that personal traits about touch may affect the user perception of the vibrotactile stimuli.

Overall, we investigated the following research question:

\begin{figure} [!t]
\centering
\includegraphics[width=2.5in]{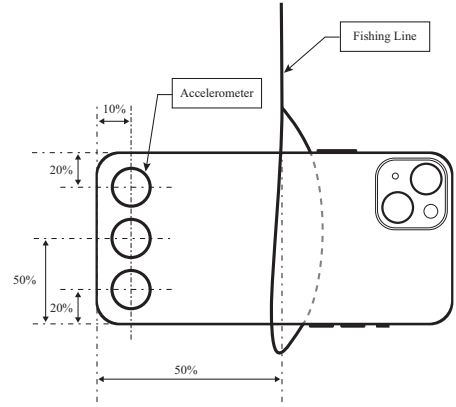}
\caption{Experimental environment to measure iPhone vibration accelerations.}
\label{fig1}
\end{figure}

\medskip
\noindent
\textbf{RQ2}: What type of differences in vibrotactile perception are caused by the physical specifications of mobile devices, the manner of operating the devices, and the personal traits of users regarding their desire for touch?

\section{Experiment}
\noindent
The aim of the experiment was measure perceptual responses to vibratory stimuli and determine the influence of various factors on their perception.
We performed magnitude estimation \cite{stevens2017psychophysics} to measure the judgments of vibratory stimuli generated from the iPhone.
To evaluate and compare how participants perceived the vibrotactile stimuli from the iPhone in the context of different sensations, we estimated the Stevens' power-law equation \cite{Stevens1957-ua} per sensation by the ordinary least square method.
The Stevens' power law is expressed as follows:

\begin{equation}
\label{eqn1}
\psi = k \phi ^{\alpha}
\end{equation}

\noindent
where $\psi$ is the perceived magnitude of the sensation evoked by a vibrotactile stimulus, $\phi$ is the amplitude of the vibration of the iPhone, and $k$ and $\alpha $ are the parameters to be estimated.

\subsection{Apparatus}\label{secApparatus}
\noindent
We used iPhone 8 or later models and iOS 15 or later iOS that used Core Haptics worked on it, which enables us to compose haptics based on the Apple Haptic and Audio Pattern (AHAP) file.

Although the precise physical specifications of different iPhone models vary, they can be broadly classified into two model groups by size: regular-size or small-size such as the iPhone SE series.
Coincidently, the proportion of iPhone ownership of each of these groups in Japan is approximately 50\% \cite{MMDLabo}; hence, we classified iPhone 8 or later models into these two model groups to increase the sample size for the experiment.
For this classification, we set iPhone 14 and iPhone SE 2nd generation as the sample models in the regular-sized model group and the small-sized model group, respectively, considering their availability to evaluate the vibratory stimuli.
We classified individual iPhone models into the closest group in terms of four physical specifications: height, width, thickness, and weight. The classification results are listed in Table \ref{tab0}.

We developed a mock application implementing the magnitude estimation.
It enabled participants to compare vibrations and evaluate the magnitude of perceived sensation on the screen of the iPhone.

\begin{table}
\begin{center}
\begin{threeparttable}
\caption{iPhone model groups}
\label{tab0}
\begin{tabular}{llcccc}
\toprule
            &           & width\tnote{a} & hight\tnote{a} & thickness\tnote{a}  & weight\tnote{a} \\
model group & model name\tnote{b} & {[}mm{]}      & {[}mm{]}        & {[}mm{]}  & {[}g{]}   \\         
\midrule
Regular     & 8 Plus     & 78.1           & 158.4          & 7.5         & 202            \\
            & X Global   & 70.9           & 143.6          & 7.7         & 174            \\
            & XS         & 70.9           & 143.6          & 7.7         & 177            \\
            & XS Max     & 77.4           & 157.5          & 7.7         & 208            \\
            & XR         & 75.7           & 150.9          & 8.3         & 194            \\
            & 11         & 75.7           & 150.9          & 8.3         & 194            \\
            & 11 Pro     & 71.4           & 144.0          & 8.1         & 188            \\
            & 11 Pro Max & 77.8           & 158.0          & 8.1         & 226            \\
            & 12         & 71.5           & 146.7          & 7.4         & 162            \\
            & 12 Pro     & 71.5           & 146.7          & 7.4         & 187            \\
            & 12 Pro Max & 78.1           & 160.8          & 7.4         & 226            \\
            & 13         & 71.5           & 146.7          & 7.65        & 173            \\
            & 13 Pro     & 71.5           & 146.7          & 7.65        & 203            \\
            & 13 Pro Max & 78.1           & 160.8          & 7.65        & 238            \\
            & \textbf{14}  & \textbf{71.5}  & \textbf{146.7}  & \textbf{7.8}  & \textbf{172}      \\
            & 14 Plus    & 78.1           & 160.8          & 7.8         & 203            \\
            & 14 Pro     & 71.5           & 147.5          & 7.85        & 206            \\
            & 14 Pro Max & 77.6           & 160.7          & 7.85        & 240            \\
Small       & 8          & 67.3           & 138.4          & 7.3         & 148            \\
            & \textbf{SE 2nd Gen} & \textbf{67.3}  & \textbf{138.4}  & \textbf{7.3}  & \textbf{148}  \\
            & 12 Mini    & 64.2           & 131.5          & 7.4         & 133            \\
            & 13 Mini    & 64.2           & 131.5          & 7.65        & 140            \\
            & SE 3rd Gen & 67.3           & 138.4          & 7.3         & 144            \\
\bottomrule
\end{tabular}

\smallskip
\scriptsize
\begin{tablenotes}
\RaggedRight
\item[a] Figures were taken from \cite{model-spec}.
\item[b] The iPhone specifications highlighted in bold font were used as the reference values for classification into each group.
\end{tablenotes}

\end{threeparttable}
\end{center}
\end{table}

\subsection{Stimuli}
\noindent
We prepared eight single vibratory pulses of 1,000 ms.
These vibratory stimuli were controlled using parameters named hapticIntensity \cite{ahapfile},\cite{hapticparm} in the AHAP file; the parameters were increased from 0.3 to 1.0 in steps of 0.1.
We set the vibration such that the hapticIntensity parameter was 0.6 as the reference stimulus in the magnitude estimation.

We measured the vibration accelerations of these stimuli in three places: lower right corner, lower center, and lower left corner, as shown in Fig. \ref{fig1}.
iPhones were held suspended in the middle with a fishing line, and an accelerometer (Ono Sokki; model NP-3414) was attached to the back sides via a magnetic base (Ono Sokki; model NP-0102) and wax.
Communication between the accelerometer and an FFT analyzer (ACO Co., Ltd.; model SpectraPLUS-SC) on Windows 10 was enabled by using a data acquisition board (ACO Co., Ltd.; model SpectraDAQ-200).
The sampling rate for measurement was 24 kHz.

\begin{figure} [!b]
\centering
\includegraphics[width=3.4in]{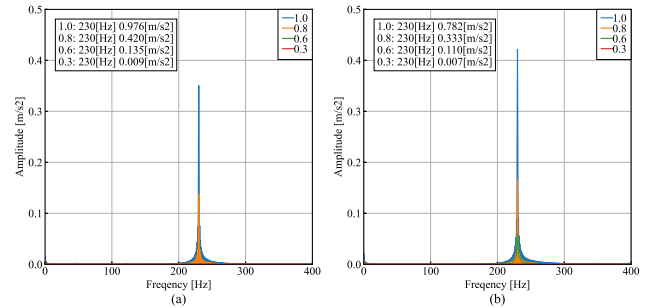}
\caption{Amplitudes of vibration controlled by hapticIntensity parameters in the AHAP file.
    (a) Amplitudes at the bottom center of iPhone SE 2nd generation and (b) at the bottom right corner of iPhone 14.
    Both devices vibrate at 230 Hz.}
\label{fig2}
\end{figure}

\begin{figure} [!b]
\centering
\includegraphics[width=3.4in]{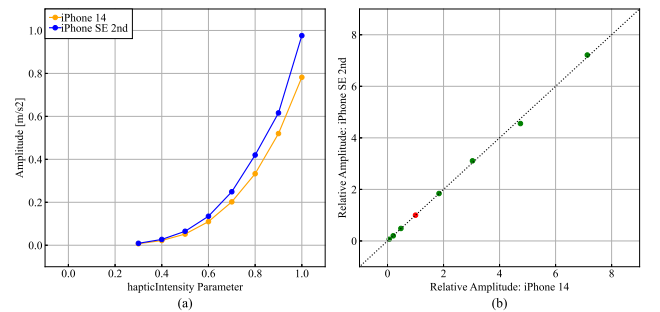}
\caption{(a) Amplitudes are nonlinear with respect to the hapticIntensity parameter.
    (b) Relative amplitudes with respect to the reference stimulus (hapticIntensity = 0.6, red dot), plotted near the dotted line corresponding to equal relative amplitudes of iPhone 14 and iPhone SE 2nd generation.}
\label{fig3}
\end{figure}

Fig. \ref{fig2} shows the maximum amplitudes of the hapticIntenstiy parameter among three measurement locations.
All vibratory stimuli were observed at 230 Hz.
We found that the amplitudes were nonlinear with respect to hapticIntensity and varied among models; meanwhile, the relative amplitudes based on the reference stimulus in the magnitude estimation were approximately the same in Fig. \ref{fig3}.

\subsection{Participants}
\noindent
We recruited participants over 18 years old with iPhone 8 or later models via Yahoo! Cloudsourcing \cite{yahoo}.
The experiment took 20 min on average, and we offered 300 yen in compensation to each participant.
Since anonymized data are handled in this study, our institutions do not require ethical review.

\subsection{Experimental procedure}
\noindent
The experimental procedure had four phases: 1) briefing and experimental condition assignment, 2) instruction, 3) experiment, and 4) debriefing.

\subsubsection{Briefing and experimental condition assignment}
At the beginning of this phase, we conducted a briefing session to explain the experiment to the participants and obtained their consent.
Then, we inquired about the iPhone models they used and whether they had accessories such as iPhone covers.

After gathering this information, we asked them to install the mock application and remove all accessories before operating the smartphone.
During the installation process, participants were randomly assigned to one of the experimental conditions for the vibrotactile sensations: intensity or weight.

\subsubsection{Instruction}
\begin{figure} 
\centering
\includegraphics[width=3.4in]{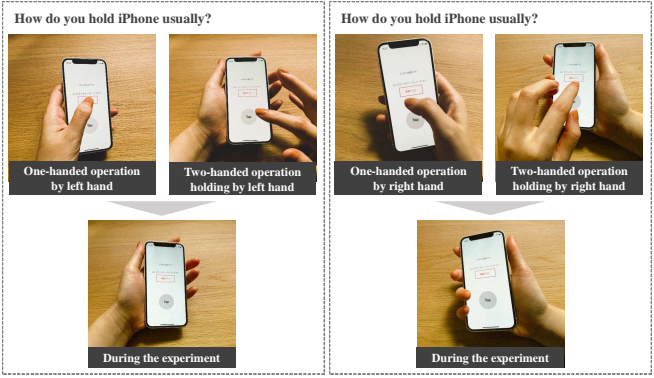}
\caption{Instructions on holding the iPhone during the experiment.}
\label{fig4}
\end{figure}

We started this phase by instructing the participants how to hold their iPhone during the experiment, as shown in Fig. \ref{fig4}.
Thereafter, we explained the base protocol for evaluating sensational magnitudes of experimental stimuli.

\begin{figure} 
\centering
\includegraphics[width=3.4in]{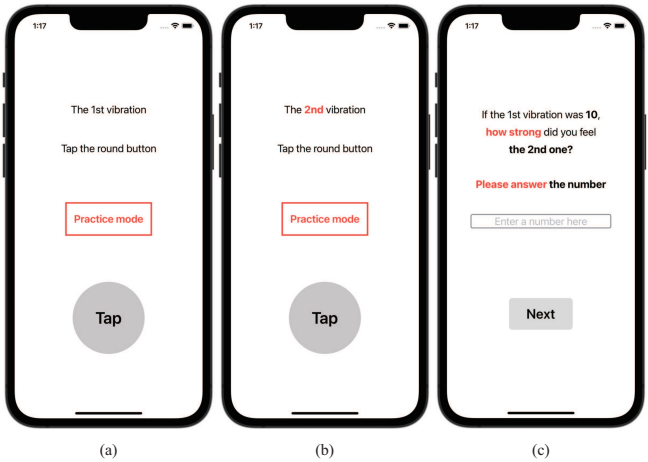}
\caption{Transition of the application screen: the base protocol for measuring levels of perceived sensation based on the magnitude estimation method.}
\label{fig5}
\end{figure}

The mock application first presented a screen, as shown in Fig. \ref{fig5}(a).
This screen called up a reference vibration when tapped and prompted the participants to perceive the vibration.
Then, the next screen, as shown in Fig. \ref{fig5}(b), was displayed; the second screen called up the experimental vibration and asked participants to perceive it.
At the end of the base protocol, the third screen was displayed, and here, participants were asked to evaluate the experimental vibration in comparison to the reference stimulus using numbers, as shown in Fig. \ref{fig5}(c).

We instructed that the base sensation magnitude of the reference stimulus was regarded as 10. Hence, we asked participants to provide the value ``20'' if they felt the magnitude of the experimental stimulus was twice the reference stimulus.
The participants assigned to the intensity (weight) condition were asked to rate the subjective intensity (weight) of the vibration as a vibrotactile sensation.

We prepared a practice mode to help the participants understand this protocol and operate the application as per their wish before the experiment.

\subsubsection{Experiment}
We prepared an experimental set comprising eight experimental stimuli, and participants randomly evaluated each stimulus in the set according to the basic protocol.
In the experiment, we offered five experimental sets repeatedly; the participants provided responses about their vibrotactile sensations 40 times.

In a postexperiment survey, we asked participants about the way in which they held the iPhone (Fig. \ref{fig4}) during the experiment.
We also asked 12 questions to measure the NFT scale \cite{Peck2003-bi}.
At the end of the experiment phase, we collected the demographics (age, sex, occupation) of the participants and information about their dominant hand and manner of holding the iPhone during daily use.

\subsubsection{Debriefing}
After measurement using the mock application, we confirmed whether the participants removed all smartphone accessories during the experiment, without imposing any penalty for the use of such accessories.

The experiment automatically ended if a participant spent more than 60 min.
And, the mock application proceeded from the instruction stage only if a participant could perceive the vibrotactile stimuli and passed the vibration check correctly.

\subsection{Sanity check}
\noindent
We presented the reference stimulus as one of the eight experimental stimuli for the instructional manipulation check.
We instructed the participants that to set the sensation magnitude of the reference stimulus as 10 (See Fig. \ref{fig5}(c)) during simulation.
Hence, we excluded all answers of participants who responded with outliers (such as zero) with respect to this reference stimulus as dishonest.
We also excluded participants whose variation among the magnitudes per hapticIntenstiy parameter was beyond three standard deviations of the mean as unreliable.
Finally, to remove possible data noises, we excluded participants who answered that they did not remove accessories during the debriefing session of the experiment. 

Consequently, the data from 167 participants were used in the analysis. The participants included 89 men aged 19–73 ($M$ =  42.7) and 78 women aged 19–56 ($M$ = 37.2).
Eighty-nine of the participants (49 men and 40 women) were assigned to the intensity condition; the remaining 78 participants were assigned to the weight condition.

\section{Results}

\begin{figure} 
\centering
\includegraphics[width=2.5in]{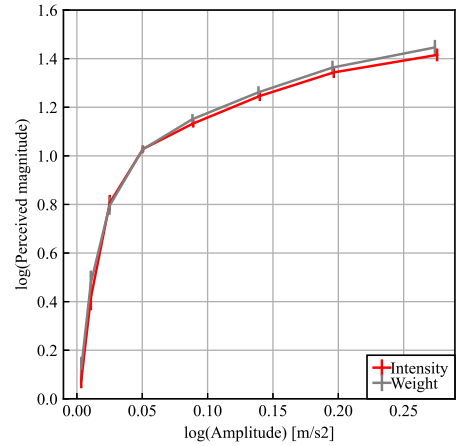}
\caption{Perceived magnitude of the vibrotactile sensation (intensity versus weight).}
\label{fig6}
\end{figure}

\subsection{RQ1}
\noindent
Fig. \ref{fig6} shows how participants perceived the magnitude of vibrotactile stimuli in different sensations.
The estimated exponent $\alpha$ of the power function was 0.65 ($p <.01$), which is fairly close to the exponent of vibration with an amplitude of 250 Hz on a finger ($\alpha = 0.60$) as evaluated in a laboratory experiment \cite{stevens2017psychophysics}.
The result showed that the vibrotactile perceptions perceived by the participants in this in-the-wild experiment were close to those perceived under laboratory conditions.

The parallelism test showed that the constant $k$ of the power-law equation varied significantly between intensity and weight conditions ($F(3, 5) = 32.201$, $p < .01$, $\eta^2 = 0.00480$).
Meanwhile, the exponent $\alpha$ showed different trends but the difference between the two conditions was insignificant ($F(3, 8) = 3.5695$, $p = 0.0589$, $\eta^2 = 0.000534$).
The results showed that the power laws governing the perceptual transformation of vibration for intensity and weight were statistically equivalent.
The context used in the perceptual transformation affected the scaling by a constant $k$ that multiplied the original power-law relation; however, this effect size $\eta^2 $ was very small. 

\subsection{Moderator variables for RQ2}
\noindent
To address RQ2, we investigated how the factors in our daily lives, such as the physical specifications of mobile devices, the manner of operating the devices, and the personal traits of users regarding their desire for touch, caused differences in the vibrotactile sensations.
We considered three factors as moderator variables for $k$ and $\alpha$ in (\ref{eqn1}): the iPhone model group as the proxy of the physical specifications of devices, the hand dominance as characteristics of the hand holding the iPhone, and the dichotomous NFT scale to assess individual differences in the process of haptic information.
We estimated the power-law equation (\ref{eqn1}) per combination of vibrotactile sensations and these factors, and subsequently, we performed multiple comparisons of the parallelism tests.

The iPhone model group was a dichotomous variable: regular-size or small-size. We obtained the model information of the iPhone that accessed the mock application during the experiment.
Using the same rule as that given in Section \ref{secApparatus}, we classified them into one of the two groups.
In all, the iPhones of 78 participants were classified into the regular-size group and those of 89 participants were classified into the small-size group.

The variable ``hand dominance'' indicated the characteristics of the hand when holding the iPhone.
Results of the postexperiment survey assigned the label ``Dominant'' to 75 participants whose dominant hand was the same as the hand that primarily supported the iPhone and ``Nondominant'' to 92 participants whose dominant hand was not used to hold the device.

The NFT scale was calculated based on responses to the 12-item questions asked in a postexperiment survey.
The entire range of the scale was from -23 to 32 in the sample, and the reliability (Cronbach's $\alpha $) of calculated NFT was .90.
As in the original paper \cite{Peck2003-bi}, a median split determined high and low NFT values: 82 subjects scoring greater than the median of 6 categorized as high in the dichotomous NFT, and those scoring less than 6 were categorized as low.

As a balance check, we confirmed that no significant sample unbalance existed between the vibrotactile sensations and the iPhone model group ($\chi ^2(1) = 0.238$, $p =.626$), hand dominance ($\chi ^2(1) = 1.579$, $p =.209$), and the NFT scale ($\chi ^2(1) = 2.127$, $p =.145$).
The chi-square test of independence showed no significant relationship among factors: the model group versus the hand dominance ($\chi ^2(1) = 0.000$, $p =.993$) and versus the NFT scale ($\chi ^2(1) = 0.509$, $p =.476$), and the hand dominance versus the NFT scale ($\chi ^2(1) = 0.774$, $p =.379$).

\subsection{RQ2}
\subsubsection{iPhone model}
\begin{figure} 
\centering
\includegraphics[width=2.5in]{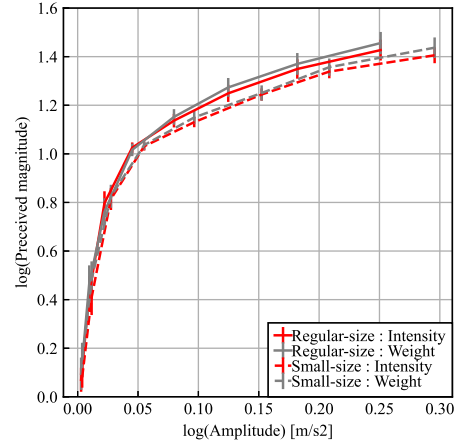}
\caption{Perceived magnitude of vibrotactile sensation by the iPhone model group (regular-size versus small-size) and sensation (intensity versus weight).}
\label{fig7}
\end{figure}

Fig. \ref{fig7} shows the perceived magnitude of vibrotactile stimuli for different combinations of sensations and iPhone model groups.
The estimated exponent $\alpha$ of the power function ranged from 0.646 to 0.664 (all $\alpha$ values had $p <.01$) among the combinations.
The parallelism test showed that the constants of the power-law equations $k$ ($F(3, 5) = 71.732$, $p <.01$, $\eta^2 = 0.0312$) and the exponents $\alpha$ ($F(3, 8) = 3.5695$, $p <.01$, $\eta^2 = 0.00264$) varied significantly among the combinations.
However, according to the effect sizes $\eta^2$, the physical specifications of the devices primarily affected the differences in the constants, and the impact was medium.

\begin{table}
\begin{center}
\begin{threeparttable}
\caption{Multiple comparison of parallel line analysis:\\ iPhone model group - sensation}
\label{tab1}
\begin{tabular}{lllll}
\toprule
 & \multicolumn{1}{l}{} & \multicolumn{1}{l}{} & \multicolumn{2}{c}{Effect size ($\eta^2$) \tnote{a,b}} \\
\multicolumn{3}{c}{Combination\tnote{c}}      & Constant ($k$) & Exponent ($\alpha$)       \\
\midrule
\multicolumn{3}{l}{(i) between models}         &                       &                \\
 & Regular-I               & Small-I                 & $\textbf{.035} ^{**}$ & $.000 ^{ }$\\
 & Regular-W               & Small-W                 & $\textbf{.018} ^{**}$ & $.004 ^{**}$ \\
\multicolumn{3}{l}{(ii) between sensations}    &                       &                \\
 & Regular-I               & Regular-W                 & $.002 ^{ }$           & $.000 ^{ }$ \\
 & Small-I                 & Small-W                 & $.007 ^{**}$          & $.003 ^{**}$ \\
\multicolumn{3}{l}{(iii) interaction}          &                       &                \\
 & Regular-I               & Small-W                 & $\textbf{.010} ^{**}$ & $.004 ^{**}$ \\
 & Regular-W               & Small-I                 & $\textbf{.048} ^{**}$ & $.000 ^{ }$ \\               
\bottomrule
\end{tabular}

\smallskip
\scriptsize
\begin{tablenotes}
\RaggedRight
\item[a] ** $p <.01$; * $p <.05$;
\item[b] Effect size values in bold font are approximately equal to or greater than ``small'' ($\eta^2$ = 0.01).
\item[c] Abbreviations are as follows:\\
         Regular-I: Regular-sized models - Intensity\\
         Regular-W: Regular-sized models - Weight\\
         Small-I: Small-sized models - Intensity\\
         Small-W: Small-sized models - Weight
\end{tablenotes}

\end{threeparttable}
\end{center}
\end{table}

Table \ref{tab1} lists the result of multiple comparisons of the parallelism test with the Bonferroni adjustment to investigate which pair of the combination was significantly different.
We considered not only the statistical significance but also the effect sizes that were at least approximately equal to or greater than ``small'' ($\eta^2$ = 0.01) to determine the factor that meaningfully affected the vibrotactile perceptions.
According to Table \ref{tab1} (ii), there is no statistical difference in the perceived magnitudes of intensity and weight within the same model group; if there was any difference, the effect sizes were tiny.
Meanwhile, from Table \ref{tab1} (i) and (iii), the difference in the model group primarily affected the level of the perceived magnitudes of both intensity and weight: more extensive sensation magnitudes were obtained from the larger and heavier devices for the same stimuli.

\smallskip
\subsubsection{Hand dominance}
Regardless of whether participants were right-handed or left-handed, approximately half of them operated their iPhones using their dominant hand. The remaining participants held their iPhones in their nondominant hands and operated them with the dominant hands.

\begin{figure} 
\centering
\includegraphics[width=2.5in]{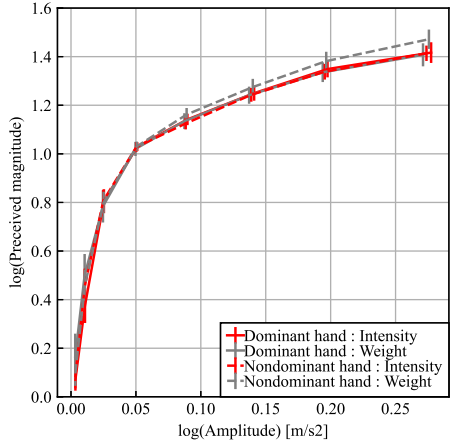}
\caption{Perceived magnitude of vibrotactile sensation when holding the iPhone in different ways (dominant versus nondominant hand) and sensation (intensity versus weight).}
\label{fig8}
\end{figure}

Fig. \ref{fig8} shows the perceived magnitude of vibrotactile stimuli in different combinations of sensations and hand dominance.
The estimated exponent $\alpha$ of the power function ranged from 0.629 to 0.669 (all $\alpha$ values were $p <.01$) among the combinations; the range of estimated exponents was wider than the iPhone model type.
The parallelism test showed that the constants of the power-law equations $k$ ($F(3, 5) = 12.186$, $p <.01$, $\eta^2 = 0.00545$) and the exponents $\alpha$ ($F(3, 8) = 17.243$, $p <.01$, $\eta^2 = 0.00769$) varied significantly among the combinations.
According to the effect size $\eta^2 $, the way people held their iPhone had only a minor influence on vibrotactile perception but had a relatively mentionable impact on the exponent than the constant.

\begin{table}
\begin{center}
\begin{threeparttable}
\caption{Multiple comparison of parallel line analysis:\\ Dominant versus Nondominant hand - sensation}
\label{tab2}
\begin{tabular}{lllll}
\toprule
 & \multicolumn{1}{l}{} & \multicolumn{1}{l}{} & \multicolumn{2}{c}{Effect size ($\eta^2$) \tnote{a,b}} \\
\multicolumn{3}{c}{Combination\tnote{c}}      & Constant ($k$) & Exponent ($\alpha$)       \\
\midrule
\multicolumn{3}{l}{(i) between hand dominance}         &                       &                \\
 & D-I                 & NonD-I                & $.001 ^{ }$           & $.003 ^{*}$\\
 & D-W                 & NonD-W                & $.000 ^{ }$           & $\textbf{.012} ^{**}$ \\
\multicolumn{3}{l}{(ii) between sensations}    &                       &                \\
 & D-I                 & D-W                   & $.005 ^{**}$          & $\textbf{.012} ^{**}$ \\
 & NonD-I              & NonD-W                & $.004 ^{**}$          & $.002 ^{*}$ \\
\multicolumn{3}{l}{(iii) interaction}          &                       &                \\
 & D-I                 & NonD-W                & $\textbf{.009} ^{**}$ & $.000 ^{ }$ \\
 & D-W                 & NonD-I                & $.002 ^{ }$           & $.004 ^{**}$ \\               
\bottomrule
\end{tabular}

\smallskip
\scriptsize
\begin{tablenotes}
\RaggedRight
\item[a] ** $p <.01$; * $p <.05$;
\item[b] Effect size values in bold font are approximately equal to or greater than ``small'' ($\eta^2$ = 0.01).
\item[c] Abbreviations are as follows:\\
         D-I: Dominant hand - Intensity; D-W: Dominant hand - Weight\\
         NonD-I: Nondominant hand - Intensity\\
         NonD-W: Nondominant hand - Weight
\end{tablenotes}

\end{threeparttable}
\end{center}
\end{table}

Table \ref{tab2} lists the result of multiple comparisons of the parallelism test with the Bonferroni adjustment. Compared to the perception of intensity in Table \ref{tab2} (ii), people perceived more weight for stronger vibrotactile stimuli on the dominant hand. Moreover, the perceived weight on the nondominant hand was more than on the dominant hand when the stimuli intensified (see Table \ref{tab2} (i)).
The estimated constant $k$ in the parallelism test for Table \ref{tab2} (iii) showed that the perceived sensational magnitude in weight on the nondominant hand was approximately 3.2\% higher than the intensity on the dominant hand.
This finding implies that the manner of holding the iPhone significantly affected the perceptual transformation of vibration.

\smallskip
\subsubsection{Need For Touch}
\begin{figure} 
\centering
\includegraphics[width=2.5in]{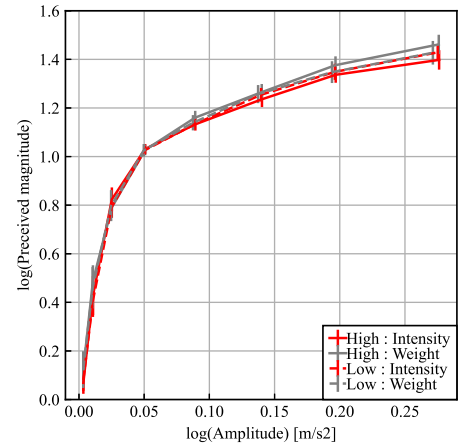}
\caption{Perceived magnitude of vibrotactile sensation depending on the Need for Touch (NFT) scale (high versus low) and sensation (intensity versus weight).}
\label{fig9}
\end{figure}

Fig. \ref{fig9} shows the perceived magnitude of vibrotactile stimuli in different combinations of sensations and dichotomous NFT scale.
The estimated exponent $\alpha$ of the power function ranged from 0.644 to 0.657 (all $\alpha$ values were $p <.01$) among the combinations; the range of the estimated exponents was the narrowest among the three factors.
The parallelism test showed that the constants of the power-law equation $k$ varied significantly among the combinations ($F(3, 5) = 11.438$, $p <.01$, $\eta^2 = 0.00511$). Meanwhile, the exponents $\alpha$ showed different but insignificant trends ($F(3, 8) = 2.472$, $p = 0.0598$, $\eta^2 = 0.00111$). 

\begin{table}
\begin{center}
\begin{threeparttable}
\caption{Multiple comparison of parallel line analysis:\\ Need For Touch - sensation}
\label{tab3}
\begin{tabular}{lllll}
\toprule
 & \multicolumn{1}{l}{} & \multicolumn{1}{l}{} & \multicolumn{2}{c}{Effect size ($\eta^2$) \tnote{a,b}} \\
\multicolumn{3}{c}{Combination\tnote{c}}      & Constant ($k$) & Exponent ($\alpha$)       \\
\midrule
\multicolumn{3}{l}{(i) between dichotomous NFT}         &                &             \\
 & High-I               & Low-I                 & $.000 ^{ }$            & $.001 ^{ }$ \\
 & High-W               & Low-W                 & $.001 ^{ }$            & $.000 ^{ }$ \\
\multicolumn{3}{l}{(ii) between sensations}     &                        &             \\
 & High-I               & High-W                & $\textbf{.008} ^{**}$  & $.000 ^{ }$ \\
 & Low-I                & Low-W                 & $.003 ^{*}$            & $.002 ^{ }$ \\
\multicolumn{3}{l}{(iii) interaction}           &                        &             \\
 & High-I               & Low-W                 & $.004 ^{**}$           & $.000 ^{ }$ \\
 & High-W               & Low-I                 & $.006 ^{**}$           & $.001 ^{ }$ \\              
\bottomrule
\end{tabular}

\smallskip
\scriptsize
\begin{tablenotes}
\RaggedRight
\item[a] ** $p < .01$; * $p < .05$;
\item[b] Effect size values in bold font are approximately equal to or greater than "small" ($\eta^2$ = 0.01).
\item[c] Abbreviations are as follows:\\
         High-I: High NFT - Intensity; High-W: High NFT - Weight\\
         Low-I: Low NFT - Intensity; Low-W: Low NFT - Weight
\end{tablenotes}

\end{threeparttable}
\end{center}
\end{table}

Table \ref{tab3} lists the result of multiple comparisons of the parallelism test with the Bonferroni adjustment.
There was no significant difference in the scaling exponents, indicating that all the power-law relations in the combination of sensations and the dichotomous NFT scores were proportionally equivalent.
The levels of perceived magnitude among the sensations and the interaction of NFT scores were significantly different in Table \ref{tab3} (ii) and (iii); meanwhile, the intensity and weight perceived by those with low NFT scores were equivalent, and their levels were moderate compared to those with high NFT scores in Fig. \ref{fig9}.
These results implied that one should consider personal traits, such as the level of haptic information processing, for vibrotactile sensational transformation.

\section{Discussion}
\noindent
Through a series of analyses, we clarified that people perceive the intensity and weight of smartphone vibrations differently.
Under the same vibration stimulus conditions, the perception of the weight of vibration stimuli tends to be higher in sensation magnitude than when they perceive the intensity.

More extensive magnitudes of weight were observed from larger and heavier devices for the same stimuli, indicating that the physical specifications of the vibration actuator affect the vibrotactile transformation to weight.
It is reasonable to assume that people unconsciously add the weight of the iPhone when translating to the perceived weight.
Our results show the importance of considering the personal traits of users, such as the level of haptic information processing, for the vibrotactile sensational transformation: People with high NFT scores are likely to distinguish different sensations from the same vibrotactile stimuli better than people with low NFT scores.
For better tactile system design, it is desirable to consider the aspects of the personal traits of users; very few studies have focused on this aspect.

The use of smartphones in daily life cannot be controlled; however, we found that the manner of holding smartphones and the hand that is used to hold it influences vibrotactile transformation. 
Unlike the previous two elements we investigated, the manner of holding smartphones affect differences in the power law to a lesser extent than the perceptual transformation of vibration among sensations. 
This result suggests some of the key issues to be considered when designing vibrotactile stimuli to perceive the weight of digital objects.
Developing user interfaces to encourage people to hold their smartphones in specific ways may be essential to maximize the effects of vibrotactile stimuli.

Limitations and future works are as follows.
The weight and size of smartphones should be separated for detailed analyses.
The size or shape of the smartphone can affect how people grasp it and how comfortable it is to use.
We simplified the user interface of the mock application and the context to create as many vibrations as possible.
The interaction of the user interface design, the context in which the vibrations are applied, and vibrotactile perception should be considered in future works.


\begin{thebibliography}{1}
\bibitem{Verrillo1969-kl}
R.~T. Verrillo, A.~J. Fraioli, and R.~L. Smith, ``Sensation magnitude of
  vibrotactile stimuli,'' \emph{Perception \& Psychophysics}, vol.~6, no.~6,
  pp. 366--372, 1969.

\bibitem{Chang2002-kh}
A.~Chang, S.~O'Modhrain, R.~Jacob, E.~Gunther, and H.~Ishii, ``{ComTouch}:
  design of a vibrotactile communication device,'' in \emph{Proc. the 4th Conf.
  Designing interactive systems: processes, practices, methods, and
  techniques}, ser. DIS '02. New York,
  NY: Association for Computing Machinery, Jun. 2002, pp. 312--320.

\bibitem{Choi2013-zs}
S.~Choi and K.~J. Kuchenbecker, ``Vibrotactile display: Perception, technology,
  and applications,'' \emph{Proc. the IEEE}, vol. 101, no.~9, pp. 2093--2104,
  2013.

\bibitem{hoggan2008crossmodal}
E.~Hoggan, T.~Kaaresoja, P.~Laitinen, and S.~Brewster, ``Crossmodal congruence:
  the look, feel and sound of touchscreen widgets,'' in \emph{Proc. the 10th
  Int. Conf. Multimodal interfaces}, ser. ICMI '08. New York, NY: Association for Computing Machinery, Oct. 2008, pp.
  157--164.

\bibitem{Rantala2020-ns}
J.~Rantala, P.~Majaranta, J.~Kangas, P.~Isokoski, D.~Akkil, O.~{\v S}pakov, and
  R.~Raisamo, ``Gaze interaction with vibrotactile feedback: Review and design
  guidelines,'' \emph{Human--Computer Interaction}, vol.~35, no.~1, pp. 1--39,
  2020.

\bibitem{Blum2019-ud}
J.~R. Blum, P.~E. Fortin, F.~Al~Taha, P.~Alirezaee, M.~Demers, A.~Weill-Duflos,
  and J.~R. Cooperstock, ``Getting your hands dirty outside the lab: A
  practical primer for conducting wearable vibrotactile haptics research,''
  \emph{IEEE Transactions on Haptics}, vol.~12, no.~3, pp. 232--246, 2019.

\bibitem{tan1999information}
H.~Z. Tan, N.~I. Durlach, C.~M. Reed, and W.~M. Rabinowitz, ``Information
  transmission with a multifinger tactual display,'' \emph{Perception \&
  Psychophysics}, vol.~61, no.~6, pp. 993--1008, 1999.

\bibitem{park2011perceptual}
G.~Park and S.~Choi, ``Perceptual space of amplitude-modulated vibrotactile
  stimuli,'' in \emph{2011 IEEE world haptics conf.} IEEE, 2011, pp. 59--64.

\bibitem{romano2011creating}
J.~M. Romano and K.~J. Kuchenbecker, ``Creating realistic virtual textures from
  contact acceleration data,'' \emph{IEEE Transactions on haptics}, vol.~5,
  no.~2, pp. 109--119, 2011.

\bibitem{Visell2014}
Y.~Visell and S.~Okamoto, ``Vibrotactile sensation and softness perception,''
  in \emph{Multisensory Softness: Perceived Compliance from Multiple Sources of
  Information}, M.~Di~Luca, Ed. London,
  UK: Springer, 2014, pp. 31--47.

\bibitem{hoggan2008investigating}
E.~Hoggan, S.~A. Brewster, and J.~Johnston, ``Investigating the effectiveness
  of tactile feedback for mobile touchscreens,'' in \emph{Proc. the SIGCHI
  Conf. Human factors in computing systems}, ser. CHI '08. New York, NY: Association for Computing Machinery, Apr.
  2008, pp. 1573--1582.

\bibitem{park2011tactile}
G.~Park, S.~Choi, K.~Hwang, S.~Kim, J.~Sa, and M.~Joung, ``Tactile effect
  design and evaluation for virtual buttons on a mobile device touchscreen,''
  in \emph{Proc. the 13th Int. Conf. Human Computer Interaction with Mobile
  Devices and Services}, ser. MobileHCI '11. New York, NY: Association for Computing Machinery, Oct. 2011, p.
  11–20.

\bibitem{Palomares2020-mw}
N.~M.~C. Palomares, G.~B. Romero, and J.~L.~A. Victor, ``Assessment of user
  interpretation on various vibration signals in mobile phones,'' in
  \emph{Advances in Neuroergonomics and Cognitive Engineering}, ser. AHFE 2019.
  Advances in Intelligent Systems and Computing, H.~Ayaz, Ed. Cham, Switzerland: Springer, 2020, pp. 500--511.

\bibitem{Park2020-db}
C.~Park, J.~Yoon, S.~Oh, and S.~Choi, ``Augmenting physical buttons with
  vibrotactile feedback for programmable feels,'' in \emph{Proc. the 33rd Annu.
  {ACM} Symp. User Interface Software and Technology}, ser. UIST '20. New York, NY: Association for Computing
  Machinery, Oct. 2020, pp. 924--937.

\bibitem{Khosravi2022-ze}
H.~Khosravi, K.~Etemad, and F.~F. Samavati, ``Mass simulation in {VR} using
  vibrotactile feedback and a co-located physically-based virtual hand,''
  \emph{Computers \& Graphics}, vol. 102, pp. 120--132, 2022.

\bibitem{Manshad2019-gv}
M.~S. Manshad, D.~Brannon, S.~A. Alharthi, and V.~Iyer, ``{Haptic-Payment}:
  Stimulating 'pain' of payment through vibration feedback in mobile devices,''
  in \emph{Proc. the 2019 {ACM} Int. Conf. Interactive Surfaces and Spaces},
  ser. ISS '19. New York, NY:
  Association for Computing Machinery, Nov. 2019, pp. 379--384.

\bibitem{core-haptics}
{Apple Inc.} {``Core Haptics: Compose and play haptic patterns to customize
  your iOS app’s haptic feedback''}. Accessed Jan. 13, 2023. [Online].
  Available: \url{https://developer.apple.com/documentation/corehaptics}


\bibitem{Abou_Chahine2022-vj}
R.~Abou~Chahine, D.~Kwon, C.~Lim, G.~Park, and H.~Seifi, ``Vibrotactile
  similarity perception in crowdsourced and lab studies,'' in \emph{Haptics:
  Science, Technology, Applications}, ser. EuroHaptics 2022. Lecture Notes in
  Computer Science, H.~Seifi, A.~M.~L. Kappers, O.~Schneider, K.~Drewing,
  C.~Pacchierotti, A.~Abbasimoshaei, G.~Huisman, and T.~A. Kern, Eds. Cham, Switzerland: Springer, 2022, pp. 255--263.

\bibitem{johansson2009coding}
R.~S. Johansson and J.~R. Flanagan, ``Coding and use of tactile signals from
  the fingertips in object manipulation tasks,'' \emph{Nature Reviews
  Neuroscience}, vol.~10, no.~5, pp. 345--359, 2009.

\bibitem{Yang2018-gy}
S.~T. Yang, S.~H. Jin, G.~Lee, S.~Y. Jeong, and J.~An, ``Dominant and
  subdominant hand exhibit different cortical activation patterns during
  tactile stimulation: An {fNIRS} study,'' in \emph{2018 6th Int. Conf.
  {Brain-Computer} Interface ({BCI})}.
  IEEE, 2018, pp. 1--3.

\bibitem{Kim2020-do}
M.-Y. Kim, H.~Kwon, T.-H. Yang, and K.~Kim, ``Vibration alert to the brain:
  Evoked and induced {MEG} responses to {High-Frequency} vibrotactile stimuli
  on the index finger of dominant and non-dominant hand,'' \emph{Frontiers in
  Human Neuroscience}, vol.~14, p. 576082, 2020.

\bibitem{Peck2003-bi}
J.~Peck and T.~L. Childers, ``Individual differences in haptic information
  processing: The ``need for touch'' scale,'' \emph{Journal of Consumer
  Research}, vol.~30, no.~3, pp. 430--442, 2003.

\bibitem{stevens2017psychophysics}
S.~S. Stevens, \emph{Psychophysics: Introduction to Its Perceptual, Neural and
  Social Prospects}. Routledge, 2017.

\bibitem{Stevens1957-ua}
------, ``On the psychophysical law,'' \emph{Psychological Review}, vol.~64,
  no.~3, pp. 153--181, 1957.

\bibitem{MMDLabo}
{Mobile Marketing Data Labo.}, ``2022-nen 5-gatsu smartphone os share chousa
  [smartphone os share survey in may 2022],'' White Paper, May 2022.

\bibitem{model-spec}
{Apple Inc.} {``Gijyutsu Shiyou [Tech Specs]''}. Accessed Jan. 13, 2023.
  [Online]. Available: \url{https://support.apple.com/ja_JP/specs}

\bibitem{ahapfile}
------. {``Representing Haptic Patterns in AHAP Files''}. Accessed Jan. 13,
  2023. [Online]. Available:
  \url{https://developer.apple.com/documentation/} \url{corehaptics/representing_haptic_patterns_in_ahap_files}

\bibitem{hapticparm}
------. {``Type Property: hapticIntensity''}. Accessed Jan. 13, 2023. [Online].
  Available:
  \url{https://developer.apple.com/documentation/corehaptics/} \url{chhapticevent/parameterid/3114595-hapticintensity}

\bibitem{yahoo}
{``Yahoo! Cloudsourcing''}. Accessed Oct. 01, 2022. [Online]. Available:
  \url{https://crowdsourcing.yahoo.co.jp/}

\end{thebibliography}

\end{document}